%% file: hicss.tex
\documentclass[10pt]{article}
\usepackage{graphicx}
\input{hicss-packages.tex}

\title{DPA Load Balancer: Load balancing for Data Parallel Actor-based systems}

\author{Ziheng Wang \\
\textit{zihengw@stanford.edu}
\And Amir Ziai \\
\textit{amirziai@stanford.edu}
\And Atem Aguer \\
\textit{atemjohn@stanford.edu}
}

\begin{document}
\maketitle

\begin{abstract}

In this project we explore ways to dynamically load balance actors in a streaming framework. This is used to address input data skew that might lead to stragglers. We continuously monitor actors' input queue lengths for load, and redistribute inputs among reducers using consistent hashing if we detect stragglers. To ensure consistent processing post-redistribution, we adopt an approach that uses input forwarding combined with a state merge step at the end of the processing. We show that this approach can greatly alleviate stragglers for skewed data.

\end{abstract}

\section{Introduction and Background}

A lot of distributed databases today follow the push-based architecture, where stateful actors push batches of data to each other in a streaming fashion [\cite{dageville2016snowflake,chen2016memsql}]. For example let's consider an aggregation query where we want to tally up the appearances of each key. In this case there would be ``mapper" actors that read chunks of the input table and feed the results to ``reducer" actors who perform the aggregation on the incoming chunks. The reducers could be hash partitioned on the key. Mappers and reducers execute simultaneously on the same cluster. Mappers push their outputs to queues on the reducers, and reducers dequeue these inputs asynchronously. This pipelined parallelism overlaps the IO-intensive mapper execution with the compute-intensive reducer computation.

This execution scheme is great when the keyspace can be evenly partitioned amongst reducers. However in real workloads the key space can be severely skewed. For example, if we are counting English words and partitioning based on the first letter, some letters (e.g. h) are a lot more common than other letters (e.g. z). This could result in some reducers having more work than other reducers, causing load imbalance in the system. We could try to sample the input ahead of time to figure out the ideal partitioning strategy, or we could try to dynamically readjust the partitioning strategy at runtime. In this work we explore the latter approach. 

Runtime load balancing can be preferable in practice if there is a high cost in obtaining accurate samples or if samples are not readily available (e.g. the input is the result of another pipeline). The latter case is especially problematic as it nullifies adaptive approaches that assume fully materialized inputs such as Hurricane and Spark's adaptive query execution [\cite{bindschaedler2018rock,zaharia2010spark}]. The state of the art in streaming load balancing today involves a coordinated global rollback to a consistent snapshot, key redistribution, and subsequent input replay [\cite{carbone2015apache}]. This can be prohibitively expensive in systems involving potentially thousands of streaming operators, who all need to be rolled back.

In this work, we develop a runtime load balancing approach without the need for coordinated global rollback. We dynamically adjust the partitioning scheme amongst the reducers leveraging consistent hashing. Our implementation handles cases where the skew can be handled by repartitioning the keyspace.

To determine when we should repartition the keyspace, we keep track of how many outstanding keys each reducer has to process as a rough proxy on the workload distribution. If we see one reducer has a lot more keys to process than the rest, we repartition the keyspace with consistent hashing to reduce its load. However, when this redistribution event occurs, there could still be inputs associated with the old partition scheme in the reducer's queue. Therefore, when a reducer sees a key it is no longer assigned to, it simply forwards it to the correct reducer. This removes the need to synchronously redistribute queue contents among reducers as effectively done in coordinated global rollback schemes.

Key redistribution could result in inputs associated with the same key being processed by different reducers throughout the program's execution. We implement a final ``state merge" step where the state of all the reducers are merged. For example, in word count, the reducer's state would be the total count of each word it has seen. If the word ``foo" is first processed by reducer A but then processed by reducer B, both A and B would have a count of foo in their state. The state merge step would simply add those counts. Other more complicated reduction functions (such as sort) might admit other custom merge functions. We discuss the implications of this design choice and explore alternatives.
\section{System Overview}
\begin{figure*}
  \centering
  \includegraphics[width=0.7\linewidth]{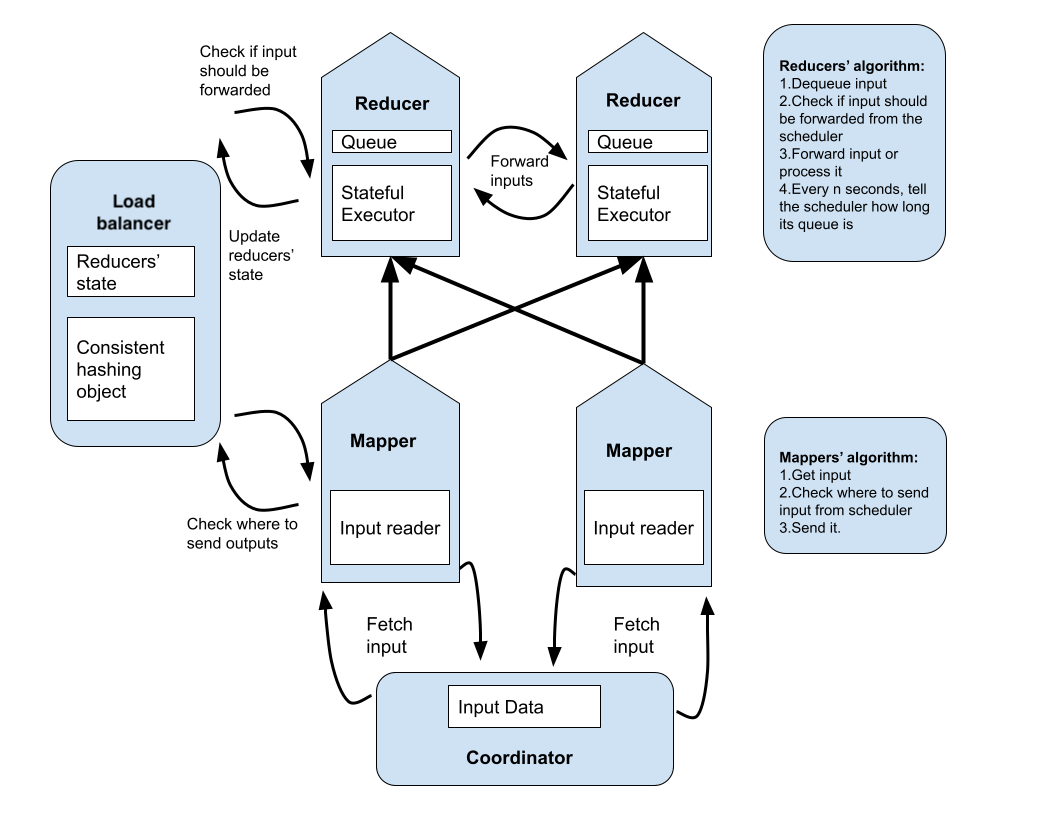}
  \caption{High level system architecture}
  \label{fig:arch}
\end{figure*}

At a high level, our system can be described as a map-reduce runtime that takes in input data, and map and reduce functions that are applied onto the input data elements. To use our system, a user provides map and reduce executors that are user-defined functions or class objects that contain arbitrary code along with input data  which are passed to a coordinator responsible for orchestrating the entire execution pipeline. The overall architecture is shown in Figure \ref{fig:arch}.

\subsection{Mappers \& Reducers}
Mappers and reducers in this streaming framework are stateful actors which get data from a source and push data to other actors. In our implementation, mappers get data from the coordinator and push data to downstream reducers. Mappers are stateless, while reducers can be stateful.  For example, for word count, the reducer maintains the count of words in dictionary and updates this dictionary upon receiving new input. Reducers also periodically  send state to the load balancer. The load balancer uses this state to determine whether a reducer is overloaded and consequently re-partition the keyspace and redistribute tasks to less overwhelmed reducers. 

\subsection{Per-Reducer Queues}
Our system leverages dynamic scheduling using per-reducer queues and therefore each reducer reads input from a specifically assigned queue. Mappers stream their outputs to reducer queues and reducers continuously consume and perform computations on these data items. The decision to use per-reducer queues stems from the fact that the interactions between our queuing subsystem and the mappers and reducers is synchronous and this leads to contention. Using per-reducer queues serves to eliminate this contention. 

\subsection{Coordinator}
The Coordinator is responsible for creating and launching the mappers and reducers, initializing the load balancer, and orchestrating the entire pipeline. It also assigns tasks to mappers and tracks the lifetimes of reducers. It determines when a reducer should shutdown after it's done processing all input data in its queue. A reducer can never stop on its own because it can still be forwarded data in the event that the other reducers are still running and one of them triggers load balancing on the load balancer. The coordinator tracks all the reducers and ensures that they shutdown once all of them are done processing the data.
\subsection{Load Balancer (Scheduler)}
The load balancer is the heart of the system. It is responsible for balancing load amongst the reducers. It achieves this through a consistent hashing object which partitions the keyspace and distributes data items among reducers. It also maintains the current load state of every reducer. This state is periodically updated by each reducer. Additionally, it exposes a method that mappers and reducers call to determine which reducer is responsible for handling a particular data item. When it detects load imbalance from the reducer states, it appropriately updates the consistent hashing object to repartition the keyspace.

\section{System Interactions}
The coordinator is responsible for instantiating and launching all system components. 

When the system starts, mapper actors fetch tasks or data items from the coordinator by means of a remote method call. The mappers apply their stateless executors to the input elements and output the result to a specific reducer queue. The specific queue is determined through our global consistent hashing object hosted in the load balancer actor. A mapper makes a remote method call to the load balancer with key of a data item and receives the index of reducer queue to output the processed data to. 

In parallel to the mapping phase, reducer actors execute in an infinite loop and continuously poll their input queues for new data. Before it processes a piece of data, it checks the load balancer to see if it is indeed assigned to this key. If it's not then the key is forwarded to the appropriate reducer. Reducers also periodically call a remote method on the load balancer to update their current load state which in our case is just the queue size. The load state can be made more fine-grained, but queue size seemed to work as a sufficient metric. 

In the event that the load balancer receives a load state from a reducer that signifies that it's overloaded, it appropriately updates the global consistent hashing object. Consequently mappers will send the skewed keys to other less overwhelmed reducers. Keys already pushed to the reducer queues according to the old partition scheme will be forwarded by the reducers.

We recognize that our design involves actors frequently checking the load balancer. This could potentially cause a centralized bottleneck in a distributed system. We believe this is acceptable as the actors are only reading, never writing. Alternatively we could use distributed consensus to ensure that all the actors agree of a certain partition scheme with the load balancer as the leader, which we discuss in the Discussion section.

\section{Load Balancing (LB)}
In this section, we describe our load balancing policy and mechanisms. We leverage consistent hashing for our key space partitioning and distribution.
\subsection{Load Balancing Policy}
Let $R$ be the number of reducers, $Q_i$ be the queue size for reducer $i$ (i.e. $R_i$), $Q_{max} = \max_{i=1}^{R}{Q_i}$ be the maximum queue length across all reducers, $x = \arg\max_{i=1}^R{Q_i}$ so that $Q_x = Q_{max}$, and $Q_{s} = \max_{i \in \{j\}_{j=1}^R - \{ x \}}{Q_i}$ be the second largest queue size.

The consistent hashing object is updated if the following predicate is true:
\begin{equation}
Q_{max} > Q_s (1 + \tau)
\end{equation}

where $\tau \ge 0$ is a configurable threshold.

The intuition behind this policy is that we want to catch cases where a single reducer has high relative load. The parameter $\tau$ allows us to control the sensitivity to skew. $\tau = 0$ would result in a very sensitive setup where no skew is tolerated, whereas a larger value would allow for tolerating more skew.

The load balancer checks this condition on a regular basis and triggers keyspace redistribution (details in the next section) if it is satisfied.

\subsection{Keyspace Redistribution}
Our keyspace redistribution strategy leverages consistent hashing [\cite{karger1997consistent}]. Each node $i$ (i.e. a reducer) gets a number of tokens $T_i$. Initially all nodes receive an equal number of tokens. These tokens are placed on a conceptual ring that represents the output space of some hash function $h$. We used Murmurhash3 [\cite{appleby2014murmurhash3}]. Token $t_{(i,j)}$ is the $j$-th token of the $i$-th node.

We represent a token using the string formatted as $r_{(i,j)}="\text{token-\{i\}-\{j\}}"$, so $h(r_{(i, j)})$ is the position of the token on the ring. Once all tokens are placed on the ring, we determine which node ID (integer) a key (string) maps to, using the following interface.

\begin{verbatim}
key_lookup(key: str) -> int    
\end{verbatim}

\begin{figure}[thb]
	\centering
	\includegraphics[width=0.7\linewidth]{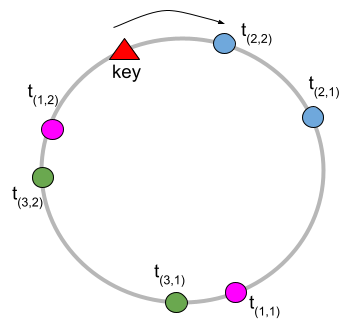}
	\caption{Key lookup in consistent hashing.}
	\label{fig:ch}
\end{figure}

We compute $h(key)$ and walk the ring clockwise starting at the hash value until we hit the first token $t_{(x, y)}$. We then return node ID $x$. Our implementation sorts all tokens by their hash values as a preprocessing step, then we can do this lookup in $O(\log{T})$ using binary search, where $T$ is the total number of tokens. Figure \ref{fig:ch} illustrate how key lookup works in an example with 3 nodes, $T_i=2$, and $T=6$.

Once we detect that a particular node is under high relative load, we trigger a redistribution of the keyspace. Ideally, we want to only target the affected node by redistributing some of the keys allocated to that node to the rest of the nodes in the system. In practice, its unclear whether this strategy would not create a bottleneck for other nodes without prior knowledge of the distribution of keys. Therefore, we introduced an interface for providing relief to the overburdened node identified with \texttt{node\_id}.

\begin{verbatim}
redistribute(node_id: int) -> None    
\end{verbatim}

We devised two strategies for keyspace redistribution.

\textbf{1- Token halving}
In this strategy we start all tokens with $N$ initial tokens where $N$ is a power of 2. When \texttt{redistribute} is called for node $i$, we remove half of the tokens owned by that node. This approach ensures that there's not a lot of disruption, other than some of the keys previously belonging to $i$ now being allocated to other nodes.

This method has a major downside: at some point a node will only have one token left and we "run out of halving".

\textbf{2- Token doubling}
We start each node with a single token. Once the method is called for node $i$, we double the number of tokens for all nodes except for $i$. The intuition is that other nodes are likely to overtake some of the keys from $i$. However, this will also result in a reshuffling of keys for the non-problematic nodes.

Neither one of these methods is guaranteed to address the load issue for node $i$. In the worst case scenario the skew is related to a single key. Even if we happen to allocate the key to another node, we would've just created a bottleneck in a new node that needs to be rebalanced later.

\section{Implementation}
Our system is implemented on top of  Ray [\cite{moritz2018ray}], a Python  framework for distributed computing. However, it can be implemented using any language (e.g. Erlang) or runtime such as Akka [\cite{akka}] that support actors. An actor in Ray is a Python object with persistent state that defines and exposes methods for interacting with that state. Actors can interact with each other by making functions calls akin to RPCs.

Each of our system components--coordinator, queues, reducers, mappers and the load balancer--are implemented as individual actors and communicate via method calls.

\section{Experiments}
In this section we describe our experimental setup. In all experiments, we fix the number of mappers and reducers to four. We use $\tau=0.2$ in all experiments.

\subsection{Metric}
We use a measure of skew (defined below) as our primary metric. Our experiments suggest that wall time is highly (inversely) correlated with this metric, so we omit it to save  space.

Additionally, we ran each experiment three times and computed the variance, which turned out to be very small, so we are omitting that as well.

\subsubsection{Skew}

Let $M_i$ be the number of messages processed by reducer $i$ and $R$ be the number of reducers. Also, let $M = \sum_{i=1}^R{M_i}$ be the total number of messages processed by all reducers.

We define $U = \lceil M / R \rceil$, which  captures the ideal scenario where messages are uniformed distributed across the reducers. We also define $W=\max_{i}{M_i}$. We define $S$ to be the normalized deviation of $W$ from ideal:
\begin{equation}
S = \dfrac{W - U}{M - U}
\end{equation}

This quantity ranges between 0 and 1. $S=0$ means that there's no skew, and $S=1$ means maximum skew (i.e. all messages are processed by a single reducer).

\subsection{Workloads}
In this section we describe the workloads that we have contrived for the purpose of studying the boundaries of our system. Workloads consist of a sequence of letters, and the goal is to produce the count of each unique letter. To ensure consistency and to speed up our experimentation, all workloads have 100 items.

\textbf{Workload 1 (WL1)} is designed so that it's skewless (i.e. $S=0$) for the halving method, but is perfectly skewed (i.e. $S=1$) for the doubling method. In other words, the initial token allocation for the halving method is such that the letters in this workload are uniformly distribution across the four nodes. For the doubling method, all items are mapped to a single reducer.

\textbf{Workload 2 (WL2)} is designed such that $S=0$ for both methods.

\textbf{Workload 3 (WL3)} is a degenerate case where the same letter is repeated 100 times (i.e. \texttt{['a', 'a', ...]}), so $S=1$ by design. 

\textbf{Workload 4 (WL4)} is heavily skewed with $S=0.8$ for the halving method and has  $S=0.49$ for the doubling method.

\textbf{Workload 5 (WL5)} is mildly skewed with $S=0.2$ for the halving method and has $S=0.55$ for the doubling method.

\subsection{Experiment 1}
We compare the two load balancing (LB) methods against the baseline (no LB) and report $S$ in Table \ref{tab:table1}. $\Delta$ is defined as the $S_{\text{No LB}} - S_{\text{With LB}}$, so a positive value signifies a reduction in $S$ when LB is applied (larger values of $\Delta$ are desirable).

We see that in cases with low skew (WL1, WL2, and WL5) these methods either don't help (i.e. $\Delta=0$) or lead to a small increase in $S$. This is due to the indeterminate nature of our distributed systems, it is possible that we trigger the LB at a point where certain reducers are behind and we don't yet have an accurate view of the load. This issue can be alleviated by choosing a larger LB threshold. However, the trade-off is that a larger threshold could lead to triggering LB past the point of it being effective.

Also, note that when initial skew is high (WL1, WL3, WL4, and WL5), we can sometimes achieve a substantial decrease in skew. In particular, the doubling method appears to be more successful in all such cases. This method is redistributing the keys more aggressively, which is more effective at distributing the load uniformly relative to surgically allocating keys from a single reducer to the other nodes (i.e. the halving method).

\begin{table}[]
\centering
\begin{tabular}{lllll}
Workload & Method   & No LB & With LB & $\Delta$ \\ \hline
WL1      & Halving  & 0.00  & 0.08    & -0.08       \\
         & Doubling & 1.00  & 0.20    & 0.80        \\ \hline
WL2      & Halving  & 0.00  & 0.00    & 0.00        \\
         & Doubling & 0.00  & 0.08    & -0.08       \\ \hline
WL3      & Halving  & 1.00  & 1.00    & 0.00        \\
         & Doubling & 1.00  & 0.75    & 0.25        \\ \hline
WL4      & Halving  & 0.80  & 0.52    & 0.28        \\
         & Doubling & 0.49  & 0.11    & 0.38        \\ \hline
WL5      & Halving  & 0.20  & 0.20    & 0.00        \\
         & Doubling & 0.55  & 0.12    & 0.43       
\end{tabular}
\caption{Experiment 1 results.}
\label{tab:table1}
\end{table}

\begin{figure*}[h]
	\centering
	\includegraphics[width=\textwidth]{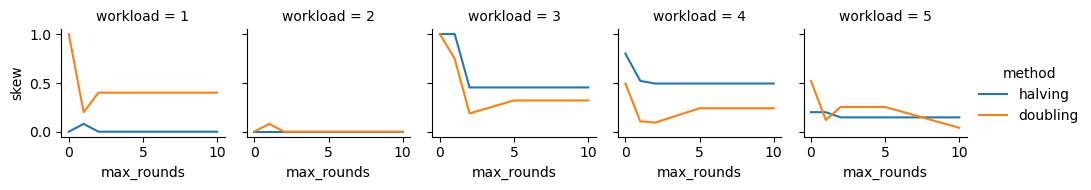}
	\caption{Experiment 2 results.}
	\label{fig:exp2}
\end{figure*}

\subsection{Experiment 2}
In this experiment we study the effects of allowing multiple rounds of LB, whereas we only allowed up to and including one round in the previous experiment.

This may be useful for two reasons: 1- as we alluded to in the previous section, sometimes LB is done prematurely and subsequent rounds can help to remedy such situations, and 2- there's no guarantee that modifying tokens will lead to the desired effects upon the first LB. It is possible that keys are remapped to the same overloaded reducer after an initial round of LB, and a subsequent round could lead to a remapping of those keys to other reducers.

Figure \ref{fig:exp2} depicts the change in skew as a function of the maximum rounds. Note that this is the maximum allowable number of rounds per reducer, so each reducer can trigger LB up to and including this many times.

We see that the additional rounds help for at least one of the methods in all workloads. WL1 and WL2 are examples of cases where both methods introduce some skew skew after the first round, which they can recover from in round 2.

Also, note that the additional rounds never hurt the halving method, but can be detrimental to doubling. This is due to the fact that reshuffling the keys many times is more likely to result in introducing skew (since many new tokens are introduced) in the doubling method vs. halving.



\section{Discussion and Future Work}
In this project, we showed that our load balancing scheme can effectively alleviate straggler effects for compute-heavy workloads by reducing the amount of items the straggler has to process. In the current implementation we simply redistributed keys among the existing reducers. However in principle our scheme can easily be extended to add new reducers on new machines. They can simply claim tokens in the consistent hashing scheme, and our forwarding mechanism will forward inputs to these new reducers appropriately. Their state has to be  merged with the state of all the existing reducers at the end. 

In our implementation, inputs associated with the same key could potentially be processed by multiple reducers. Thus the ``state" associated with this key can also be distributed on multiple reducers and might have to be merged at the end. This merge step might be expensive, and might not always be possible for non-commutative or non-associative reduction functions.

An alternative design could involve state forwarding. If at time $t$ in the program's execution, the load balancer decides to assign key $k$ from reducer A to reducer B, before reducer A forwards any input in its queue associated with key $k$ to reducer B, it could forward to reducer B all its state associated with the key $k$. In this state forwarding scheme, logically, the state associated with a particular key is always resident on a single reducer and therefore no state merging is needed at the end. In addition, this state forwarding could ideally be done asynchronously and impose no latency penalty.

A key challenge with this state forwarding approach is that the mappers can be aware of the key redistribution instantly at time $t$ and decide to forward inputs associated with key $k$ immediately to reducer $B$, before reducer $A$ has had a chance to forward $B$ the state. Reducer $B$ then would be faced with inputs in its queue with no corresponding state. Depending on reducer B's execution semantics it might decide to throw away such inputs (e.g. hash join not matching on build table), leading to incorrect execution behavior. 

One solution would be to never update where the mappers send their outputs. Thus all future inputs associated with $k$ will still arrive at reducer A, and reducer A can simply enforce the rule that it forwards the state associated with any key $k$ before forwarding any inputs associated with key $k$. However, this leads to unnecessary network traffic and redundant forwarding. 

A better solution would be this algorithm: the load balancer keeps a read-only version of the ideal keyspace partitioning which it can update at its leisure. The updates are assumed to be very infrequent and atomic. All reducers agree to this partitioning scheme through consensus. Importantly, every reducer can only exist in two states: synchronizing (undefined behavior) or synchronized (must agree on one partitioning scheme). The reducer cannot perform any other actions while it is synchronizing. It's free to process data only in the synchronized state. 

The processing is thus broken down into stages where all reducers are synchronized (i.e. agree on the partitioning scheme). The stage is broken down into two substages. In the first substage all reducers exchange state as specified by the partitioning scheme. In this substage, reducers cannot forward any data because they are not certain that the forward destination has the state to handle said data. Any data that need to be forwarded gets put back into the queue. After all state reshuffling is complete, the reducers enter the second substage where they are free to forward data as they wish. Note the data they forward might have arrived several stages back from an old partition scheme long forgotten. But the first substage guarantees that if the reducer follows the current partitioning to forward this data, the state to process that data is resident on the forwarding destination.

We plan to implement this algorithm in a research streaming system called Quokka in the coming months: https://github.com/marsupialtail/quokka.









\printbibliography

\end{document}

%% file: hicss-packages.tex
\usepackage[letterpaper]{geometry}
\usepackage{hicss}
\usepackage{times}
\usepackage[none]{hyphenat}
\usepackage{url}
\usepackage{latexsym}
\usepackage{amsmath}
\usepackage{indentfirst}
\usepackage{graphicx}
\graphicspath{{images/}}
\usepackage[
    style=apa,
  ]{biblatex}